# The Redundancy Matrix as a Performance Indicator for Structural Assessment

David Forster[1] and Malte von Scheven[1]

[1] University of Stuttgart, Institute for Structural Mechanics, Stuttgart, Germany

**Abstract**
The degree of static indeterminacy and its spatial distribution characterize load-bearing structures independent of a specific load case. The redundancy matrix stores the distribution of the static indeterminacy on its main diagonal, and thereby offers the possibility to use this property for the assessment of structures. It is especially suitable to be used in early planning stages for design exploration. In this paper, performance indicators with respect to robustness and assemblability are derived from the redundancy matrix. For each of the performance indicators, a detailed matrix-based derivation is given and the application is showcased with various truss examples.

## 1 Introduction

### 1.1 Motivation

In civil engineering, several requirements must be satisfied when designing a building. Besides aesthetic and sustainability aspects, a key aspect of the design is structural safety, meaning that the structure withstands external forces such as wind and dead load but also temperature changes and exceptional influences like vehicle impact. The national building codes are mainly restricting stresses for the ultimate limit state and displacements for the serviceability limit state, taking into account different load cases and safety factors depending on the probability of their respective occurrence (*DIN EN 1991-1-7 2010*; *ASCE 2022*). Those concepts are well-defined and known to structural engineers.

In contrast, the notions of redundancy and robustness, as well as the degree of static indeterminacy and the distribution of internal constraint are only vaguely touched in building codes. Especially the quantification of these structural performance indicators is not specified. The redundancy matrix and thus the distribution of the degree of static indeterminacy in the structure expands the possibilities for structural engineers to assess also these aspects of structural design on a quantitative basis.

Dealing with robustness, according to the German building codes, collapse must be prevented and the effect of damage and its cause must be somewhat proportional (Ellingwood and Dusenberry 2005). This means, for example, that a small event must not lead to an overall collapse of the structure.

Another aspect of structural assessment, which is not covered at all in building codes, is the assembly process and the interplay between prefabrication and on-site manufacturing. Geometric imperfections can cause an initial stress-state in the assembled structure and by this influence the load-bearing behavior. Adjusting the design or assembly sequence, the assemblability of structures can be improved and the initial stresses caused by imperfections can be kept minimal.

In this paper, quantitative design criteria for the robustness and the assemblability of structures based on the redundancy matrix are proposed. When dealing with robustness and structural assembly, the redundancy matrix serves as a suitable measure, since it is a measure for the internal constraint and independent of specific load cases. This is especially important in very early design stages, where various design options with different topologies, cross-sections and geometries need to be assessed without explicit knowledge about load cases and governing load combinations.

## 1.2 State of the Art

The concept of the redundancy matrix and the related distributed static indeterminacy, as it is used in the present paper, was proposed in the group of Klaus Linkwitz in the context of geodesic and structural mechanics research (Linkwitz 1961). Based upon this work, Bahndorf (1991) and Ströbel (1997) describe the redundancy matrix as an idempotent matrix, quantifying the spatial distribution of the degree of static indeterminacy on its main diagonal, referred to as redundancy. Ströbel and Singer (2008) use this information to quantify the sensitivity towards imperfection of a structural system. In von Scheven et al. (2021), the matrix-based derivation of the redundancy matrix is summarized for trusses and plane beams and extended to continuous systems. Applications in the field of adaptive structures are presented in Wagner et al. (2018) and Geiger et al. (2020), using the redundancy matrix for actuator placement to compensate for external loads with force or displacement manipulation within the structure. Forster et al. (2023) give a brief overview of the concept of using the redundancy for the design of structures and describe the calculation for three-dimensional frame structures. In Gil Pérez et al. (2023), the redundancy matrix is used to assess robotically assembled carbon fiber composite structures with a focus on capturing deviations stemming from geometric imperfections due to the manufacturing process.

The concept of redundancy is of course closely related to the notion of robustness. But there exists a much larger variety of definitions of robustness in the literature. Not all of them can be mentioned here. Only a few of these definitions of robustness make use of the redundancy matrix or the static indeterminacy. We will show later that in fact the redundancy distribution as a measure for robustness is identical to other definitions in the literature.

Khan and El Nimeiri (1983) present the idea that the factors of safety in designing structural elements should be adapted according to the member's importance. Based on the probabilistic approach of the building codes, the authors present a reliability index, which shifts the normal distribution curve for resistance depending on whether the member is of high importance for the load transfer or highly redundant.

Frangopol and Curley (1987) are taking into account brittle, ductile, and hardening behavior of the structural elements to assess redundancy. Amongst several definitions and interpretations of structural redundancy, the authors define four different criteria. Two of those are the degree of static indeterminacy (Maxwell 1864) and the load bearing capacity of different states of the structure, which makes this measurement of robustness dependent on a specific load case scenario. Feng and Moses (1986) use an index based on the material's strength to quantify the performance and show related optimization of structures. The above-mentioned four different criteria are also used for structural optimization and the extension of redundancy for continuous structures (Frangopol and Klisinski 1989; Pandey and Barai 1997).

The contribution by Pötzl (1996) distinguishes between the different causes of damage, disregarding external influences, and points out that the majority of structural defects are due to the design, followed by wrong execution and improper use. This underlines the fact that early design stages are of utmost importance when it comes to structural safety. Therein, the term redundancy is defined as the structure's ability to provide different load paths in order to compensate for individual failure of members, adding safety to the structure beyond the requirements in building codes. The contribution also raises the question of manufacturing and therefore the assembly process as a measure for structural performance.

Describing examples of structural collapse due to missing robustness, Harte et al. (2007) propose a quantification of robustness using a score which is again dependent on a distinct external exposure. A list of measures to design robust structures includes the structure itself but also the maintenance and the used material. Baker et al. (2008) present a framework based on probabilistic risk analysis, quantifying the direct consequences of damage as well as subsequent impacts.

Kanno and Ben-Haim (2011) define a so-called strong redundancy, taking into account the spatial distribution of the static indeterminacy within the structure. Within a truss example, this strong redundancy counts the maximum number of elements that can be removed before the structure fails, without taking into account the order. By this, the method identifies critical paths and non-redundant parts within a structure. Kou et al. (2017) present examples in the context of redundancy and robustness. They introduce a recursive method to calculate the redundancy matrix for the modified system after failure of a certain structural element in order to capture progressive failure.

Another important aspect for the assessment of structures is the assembly process and the induced stress-states during on-site assembly. In construction industry, tasks on site are mainly performed manually by skilled workers, offering the opportunity to account for dynamic environmental changes and uncertainties, while at the same time assistance through automatic control to execute repetitive tasks increases (Jin et al. 2021). With an increasing digitization and automation in construction industry, as described *e.g.* by Knippers et al. (2021), effects from predefined assembly sequences and manufacturing imperfections on the performance of the structural system need to be addressed.

Many publications deal with assembly planning to reduce the amount of formwork or even achieve self-supporting structures. Kao et al. (2017) use a method based on so-called backward assembly planning (Lee 1991) to assemble shell structures with a minimum amount of formwork. Imperfections in the manufacturing process, which can impact the initial stress-state of the assembled structure, are not taken into account. Also, recent publications in the field of robotically assembled structures mainly deal with self-supporting structures that avoid scaffolding, without referring to stress-states or imperfection sensitivity of the assembly process (Parascho et al. 2020; Bruun et al. 2022). In the context of robotically aided on-site assembly, Lauer et al. (2023) present an automated process for timber cassettes that is showcased on a real construction site. Leder et al. (2019) show the automated assembly of spatial timber structures using single-axis robots and standardized timber struts.

Manufacturing imperfections and initial stresses induced during the assembly procedure are not considered in most of these publications. Since manufacturing imperfections introduce states of stress in a structure, the ultimate load-bearing capacity can be reduced by these initial stresses. Therefore, from a structural engineering point of view, it is important to either minimize the imperfections or to decrease their negative effect by a customized assembly sequence. Within this paper, the influence of manufacturing imperfections on the strain distribution of a structure is presented. Subsequently, the effect of different assembly sequences, which lead to different structural configurations, on intermediate strain distributions is shown.

## 1.3 Outline

The paper is structured as follows. Section 2 briefly introduces the theoretical fundamentals of structural mechanics, including matrix structural analysis, the definition of the redundancy matrix and its properties. In Section 3, a measure for robustness based on the redundancy matrix is derived and showcased with a 3D truss structure. Section 4 shows the assessment of a structure in regard to the assembly and the respective derivation of a quantitative measure. Section 5 summarizes the work and gives an outlook on future research.

## 2 Fundamentals of Structural Mechanics

## 2.1 Matrix Structural Analysis

In this section, relevant quantities and equations of matrix structural analysis for linear static analysis of discrete models of spatial truss and frame structures are summarized. The formulation is based on the natural mode formulation originally presented by Argyris (1964) and Argyris et al. (1964). This formulation describes the deformation of an element by decoupled strain inducing modes and rigid body modes.

Given is a discrete model consisting of $n$ degrees of freedom, $n_n$ nodes, and $n_e$ elements, each of which carries loads via $n_m$ load-carrying modes. The number of load-carrying modes is equal to the number of generalized stress resultants or generalized elastic deformations in this element and is $n_m = 1$ for plane or spatial truss elements, $n_m = 3$ for plane beam elements and $n_m = 6$ for spatial beam elements. In general, models can consist of a combination of truss and beam elements, *i.e.*, $n_m$ can vary between the elements. Therefore, the total number of load-carrying modes of all elements is introduced as $n_q$. For models consisting of only one element type $n_q = n_m n_e$.

The relation between the external loads $\mathbf{f} \in \mathbb{R}^n$ and the generalized displacements $\mathbf{d} \in \mathbb{R}^n$ is described by the three field equations static equilibrium, elastic material law and compatibility:

$$\mathbf{A}^\top \mathbf{s} = \mathbf{f}, \qquad \mathbf{s} = \mathbf{C}\mathbf{e}_{el}, \qquad -\mathbf{e}_{el} = -\mathbf{A}\mathbf{d} + \mathbf{e}_0. \tag{1}$$

$\mathbf{A}^\top \in \mathbb{R}^{n \times n_q}$ is the equilibrium matrix, $\mathbf{A} \in \mathbb{R}^{n_q \times n}$ is the compatibility matrix, and $\mathbf{C} \in \mathbb{R}^{n_q \times n_q}$ is the material matrix, which is a diagonal matrix with positive entries. The vector $\mathbf{s} \in \mathbb{R}^{n_q}$ represents the generalized stress resultants of all elements, $\mathbf{e}_{el} \in \mathbb{R}^{n_q}$ represents the corresponding generalized elastic deformations and $\mathbf{e}_0 \in \mathbb{R}^{n_q}$ represents the generalized pre-deformations.

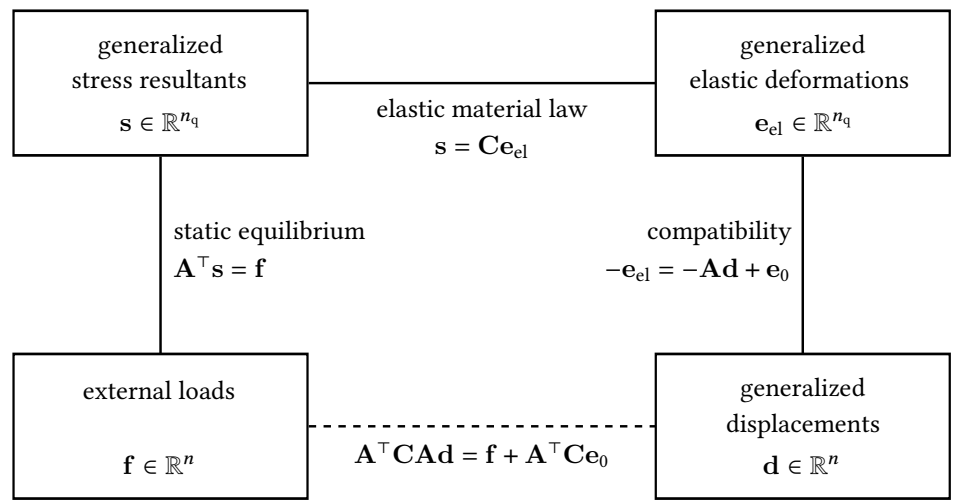

**Figure 1** Overview of relevant equations and quantities in matrix structural analysis for linear elastostatics (inspired by Tonti's diagram for elastostatic problems (Tonti 1976) and by Strang (1986)).

The diagram in Figure 1 summarizes the relevant equations and quantities in matrix structural analysis for linear elastostatics and states the equation to compute the generalized displacements $\mathbf{d}$ from the external loads $\mathbf{f}$:

$$\mathbf{K}\mathbf{d} = \mathbf{f} + \mathbf{A}^\top \mathbf{C} \mathbf{e}_0 \quad \text{with} \quad \mathbf{K} = \mathbf{A}^\top \mathbf{C} \mathbf{A}. \tag{2}$$

$\mathbf{K} \in \mathbb{R}^{n \times n}$ is called the elastic stiffness matrix. It is symmetric by definition due to the diagonality of $\mathbf{C}$.

It is assumed throughout the paper that the structures are statically indeterminate with a degree of static indeterminacy $n_s = n_q - \text{rank}(\mathbf{A}^\top)$. Furthermore, it is assumed that the structures are kinematically determinate, *i.e.*, $\text{rank}(\mathbf{A}) = n$ (Pellegrino and Calladine 1986; Pellegrino 1993), which is equivalent to $\mathbf{K}$ being regular. The latter assumption can be satisfied by properly choosing structural topology and boundary conditions. It ensures that the structures are able to equilibrate loads without pre-stress (and thus geometric stiffness effects) such that linear structural theory is applicable.

## 2.2 Definition of the Redundancy Matrix

Based on the quantities and equations of matrix structural analysis defined in the previous subsection, the concept of the redundancy matrix (Linkwitz 1961; Bahndorf 1991; Ströbel 1997; von Scheven et al. 2021) is recapitulated in the following. As state-of-the-art, the redundancy matrix is only defined for the linear setting.

The redundancy matrix is a measure of the internal constraint in a structure and is therefore independent of the external loads. Thus, $\mathbf{f} = \mathbf{0}$ is assumed. Solving Equation (2) for the generalized displacements $\mathbf{d}$ and inserting those into the compatibility Equation (1) yields a relation between the negative generalized elastic deformations $-\mathbf{e}_{\text{el}}$ and the generalized pre-deformations $\mathbf{e}_0$:

$$-\mathbf{e}_{\text{el}} = (\mathbf{I} - \mathbf{A}\mathbf{K}^{-1}\mathbf{A}^\top \mathbf{C})\mathbf{e}_0 = \mathbf{R}\mathbf{e}_0, \tag{3}$$

with the redundancy matrix $\mathbf{R} \in \mathbb{R}^{n_q \times n_q}$

$$\mathbf{R} = \mathbf{I} - \mathbf{A}\mathbf{K}^{-1}\mathbf{A}^\top \mathbf{C} \tag{4}$$

and the identity matrix $\mathbf{I} \in \mathbb{R}^{n_q \times n_q}$.

Considering Equation (3), the redundancy matrix component $R_{ik}$ maps the initial elongations imposed in element $k$ onto the negative elastic elongations in element $i$. Therefore, the redundancy matrix contains column-wise the negative generalized elastic deformations caused by a unit generalized pre-deformation in the respective element $k$. For a truss system, this corresponds to removing element $k$ from the structure and reassembling it after assigning a unit elongation. Squeezing this imperfect element into the structure will cause elastic deformations in other elements (column $k$ of the redundancy matrix). The amount by which the initial elongation in element $k$ is reduced by the surrounding structure is a measure of the constraint imposed on the element and also its redundancy in the structure. Figure 2 shows a truss structure with the redundancies in color scheme on the left and the respective redundancy matrix on the right. For a very high constraint, the resulting total deformation in element $k$ will be close to zero, the elastic deformation close to one and also the redundancy $R_{kk}$ will be close to one. In case that there are two fixed supports on either end of a truss element, see element 8 in Figure 2 and the belonging row and column in the matrix, the redundancy is one since this element is completely constrained and is not required for the load transfer. On the contrary, an element with little constraint from the surrounding structure will yield a large total deformation and a small elastic deformation and therefore a small diagonal entry and redundancy. If an element is not constrained at all by the surrounding, as it is the case for elements 6 and 7 in Figure 2, the redundancy is zero. These two elements are a statical determinate part of the structure. For elements 1 to 5, the redundancies can be seen in color scheme and read from the matrix.

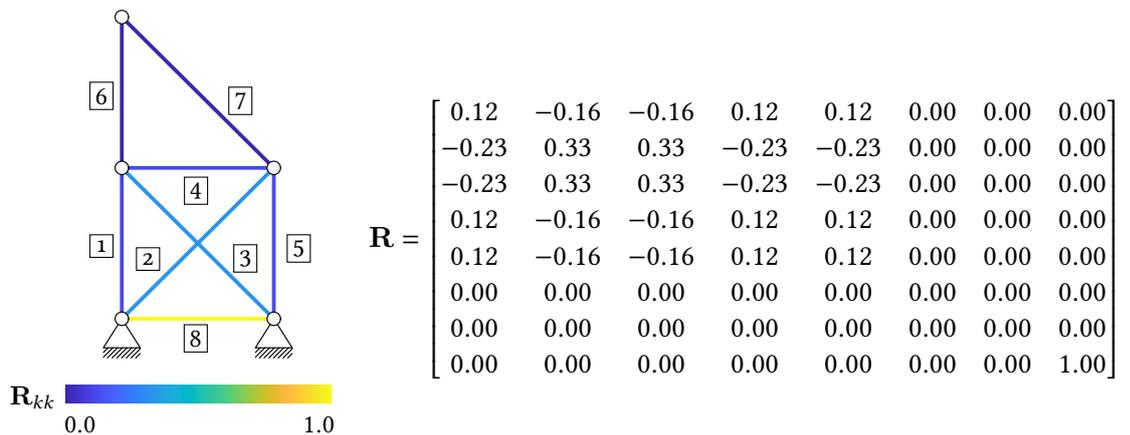

**Figure 2** Truss structure with constant cross-sections, redundancy shown in color scheme (left) and respective redundancy matrix according to Equation (4) (right).

This definition of the redundancy can be applied to all discrete structural systems, like truss systems in 2D and 3D as well as frame systems in 2D (von Scheven et al. 2021) and 3D (Forster et al. 2023; Tkachuk et al. 2023).

## 2.3 Properties of the Redundancy Matrix

The redundancy matrix $\mathbf{R}$ describes a parallel projection of initial elongations into the subspace of elastic elongations (im($\mathbf{R}$)) parallel to the subspace of compatible elongations (ker($\mathbf{R}$)).

The matrix $\mathbf{R}$ is idempotent and its trace is equal to $n_e - n_d = n_s$, $n_s$ being the total degree of static indeterminacy in the structure (von Scheven et al. 2021). As $\text{tr}(\mathbf{R}) = n_s$, the diagonal entries $R_{kk}$ of the redundancy matrix can be interpreted as the contributions of the individual elements to the total degree of static indeterminacy $n_s$ (Bahndorf 1991; Ströbel 1997). Therefore, the diagonal entries are also called distributed static indeterminacy (Eriksson and Tibert 2006; Zhou et al. 2015; Chen et al. 2018). This allows to distribute the total degree of static indeterminacy $n_s$ amongst the elements of the structure. Properties known for statically determinate or indeterminate structures can be transferred to the element level: Constraint load cases will not yield internal forces in an element with zero redundancy as this element is statically determinate and removing a statically determinate element, *i.e.* an element with zero redundancy, will lead to (partial) failure of the structure. This makes the redundancy matrix very useful for the assessment of structures with respect to robustness (reducing the impact of element failure) and assemblability (avoiding stresses due to geometrical imperfections).

The redundancy matrix $\mathbf{R}$ can also be interpreted as an influence matrix. It describes the influence of initial deformations on the elastic deformations in the structure. In some cases, it is not the influence on deformations that is important, but the influence on stresses or stress resultants. Then the elastic deformations can be directly converted into the stress resultants using the material matrix $\mathbf{C}$. The influence matrix for the stress resultants is $-\mathbf{CR}$.

Interactive design methods require fast feedback to inform designers and assist them in their decision-making process. Direct feedback on the redundancy distribution in a structure is particularly useful for topology exploration with respect to assemblability and robustness (Forster et al. 2023). But due to the inverse of the stiffness matrix and the matrix-matrix multiplications, the computational complexity for the calculation of the redundancy matrix is given by $O(n \cdot n_q^2)$. Since $n$ is typically proportional to $n_q$, the complexity scales cubically with the problem size. A more efficient computation of the redundancy matrix is proposed by Tkachuk et al. (2023). The closed-form expression is derived via a factorization of the redundancy matrix that is based on singular value decomposition.

If in a design or optimization process, a structure is iteratively examined with the help of slight adjustments, the resulting changes to the redundancy matrix can be computed via a rank one update. A generic algebraic formulation for efficiently updating the redundancy matrix (and related matrices) is presented by Krake et al. (2022). The formulations based on the Woodbury formula include various modifications like adding, removing, and exchanging elements and are applicable to truss and frame structures.

## 3 Robustness

The redundancy distribution within a structure can be used to quantify its robustness. We assume a system to be robust if the change in elastic deformations due to a given load is minimized in the event that an element fails and is therefore removed. A detailed derivation is conducted, starting from the change in stiffness due to the removal of an element up to a compact form of the calculation of the effect on the elastic deformations, see also (Bahndorf 1991). The details and the notation for the matrix calculation are based on Krake et al. (2022). Thereby, the compatibility matrix $\mathbf{A}$, the material matrix $\mathbf{C}$, and the stiffness matrix $\mathbf{K} = \mathbf{A}^\top \mathbf{C} \mathbf{A}$ refer to the initial system. The following derivation is based on one load-carrying mode $n_m$. For beam structures, the modes can be evaluated separately. The removed element is denoted as $r$, thus, the element's redundancy of the removed element is given by $R_{rr}$. The row of the compatibility matrix related to the element to be removed is described by $\mathbf{a}_r \in \mathbb{R}^{1 \times n}$ and its stiffness by $\mathbf{C}_{rr} = c_r$.

With this at hand, we can write the flexibility matrix of the modified system as the inverse of the stiffness matrix of the modified system $\tilde{\mathbf{K}}$ as

$$\tilde{\mathbf{K}}^{-1} = \left(\mathbf{A}^\top \mathbf{C} \mathbf{A} - \mathbf{a}_r^\top c_r \mathbf{a}_r\right)^{-1}. \tag{5}$$

Using the Woodbury formula (Woodbury 1950), the above equation can be rewritten as

$$\tilde{\mathbf{K}}^{-1} = \mathbf{K}^{-1} + \mathbf{K}^{-1} \mathbf{a}_r^\top c_r \left(1 - \mathbf{a}_r \mathbf{K}^{-1} \mathbf{a}_r^\top c_r\right)^{-1} \mathbf{a}_r \mathbf{K}^{-1}. \tag{6}$$

Since we want to examine the change of flexibility if an element is removed, we define

$$\Delta \mathbf{K}^{-1} = \tilde{\mathbf{K}}^{-1} - \mathbf{K}^{-1} = \mathbf{K}^{-1}\mathbf{a}_r^\top \left(c_r^{-1} - \mathbf{a}_r \mathbf{K}^{-1}\mathbf{a}_r^\top\right)^{-1} \mathbf{a}_r \mathbf{K}^{-1} \qquad (7)$$

as the change of flexibility matrix when removing element $r$.

According to Tkachuk et al. (2023), the main-diagonal entry of the redundancy matrix for the element to be removed, $R_{rr}$, can be computed as

$$R_{rr} = 1 - \mathbf{a}_r \mathbf{K}^{-1} c_r \mathbf{a}_r^\top. \qquad (8)$$

This expression can be re-written as the ratio of the redundancy of the element and the stiffness of the element as

$$\frac{R_{rr}}{c_r} = \left(c_r^{-1} - \mathbf{a}_r \mathbf{K}^{-1}\mathbf{a}_r^\top\right). \qquad (9)$$

Inserting Equation (9) into Equation (7), the change in flexibility can be expressed as

$$\Delta \mathbf{K}^{-1} = \mathbf{K}^{-1}\mathbf{a}_r^\top \frac{c_r}{R_{rr}} \mathbf{a}_r \mathbf{K}^{-1}. \qquad (10)$$

The change in displacements due to the removal of element $r$ under an arbitrary load $\mathbf{f}$ can be calculated using the change in the flexibility matrix:

$$\Delta \mathbf{d} = \Delta \mathbf{K}^{-1}\mathbf{f} = \mathbf{K}^{-1}\mathbf{a}_r^\top \frac{c_r}{R_{rr}} \mathbf{a}_r \mathbf{K}^{-1}\mathbf{f}. \qquad (11)$$

To further simplify this expression, it can be multiplied by the row of the compatibility matrix related to the removed element $\mathbf{a}_r$. This yields the change of elongation $\Delta e_r$ of the removed element, or as the element is removed, the change in distance between the corresponding nodes considering linear kinematics.

$$\Delta e_r = \mathbf{a}_r \Delta \mathbf{d} = \mathbf{a}_r \mathbf{K}^{-1}\mathbf{a}_r^\top \frac{c_r}{R_{rr}} \mathbf{a}_r \mathbf{K}^{-1}\mathbf{f}. \qquad (12)$$

Although this is only a local criterion, it describes the effect on the load-bearing behavior at the location and in the direction of the structural modification. Using Equation (8), rewritten as $1 - R_{rr} = \mathbf{a}_r \mathbf{K}^{-1} c_r \mathbf{a}_r^\top$, we can formulate the above equation as

$$\Delta e_r = \mathbf{a}_r \Delta \mathbf{d} = \frac{1-R_{rr}}{R_{rr}} \mathbf{a}_r \mathbf{K}^{-1}\mathbf{f} = \frac{1-R_{rr}}{R_{rr}} \mathbf{a}_r \mathbf{d}. \qquad (13)$$

Equation (13) shows that the change in element elongation $\Delta e_r$ caused by removing the element $r$ depends on the factor $(1-R_{rr})/R_{rr}$ and therefore on the redundancy of the removed element $R_{rr}$. The larger the redundancy of the removed element $R_{rr}$, the smaller the effect on the load-bearing behavior of the structure. This means that for a robust behavior, the redundancy of the removed element should be as large as possible.

Since robust behavior of a structure is associated with being independent of the element to fail, the redundancies of all elements need to be as large as possible. As the sum of all redundancies equals the degree of static indeterminacy and is independent of the element to be removed, a homogeneous distribution maximizes all redundancies, and thus can be used as an objective to design robust structures.

This definition of a robust structure having a homogeneous distribution of redundancy is in fact identical to other definitions in literature. The determinant of the global stiffness matrix $\det(\mathbf{K})$ is widely used to quantify robustness such that the ratio of the determinant of the modified stiffness matrix and the determinant of the initial stiffness matrix is used as a measure and maximized. Nafday (2011) denotes this ratio as the member consequence factor, used to quantify structural integrity and Starossek and Haberland (2011) use this ratio to define a stiffness-based measure of robustness.

| Element $k$ | 1 | 2 | 3 | 4 | 5 | 6 | 7 | 8 | 9 | 10 | 11 | 12 | 13 | 14 |
|---|---|---|---|---|---|---|---|---|---|---|---|---|---|---|
| $R_{kk}$ | 0.08 | 0.08 | 0.35 | 0.35 | 0.58 | 0.58 | 0.58 | 0.58 | 0.35 | 0.35 | 0.08 | 0.08 | 0.49 | 0.49 |
| $R_{kk,\text{opt}}$ | 0.36 | 0.36 | 0.36 | 0.36 | 0.36 | 0.36 | 0.36 | 0.36 | 0.36 | 0.36 | 0.36 | 0.36 | 0.36 | 0.36 |

**Table 1** Redundancies for initial configuration and optimized configuration per element.

With the help of a rank one update (Meyer 2008), the determinant of the modified stiffness matrix can be written as

$$\det(\tilde{\mathbf{K}}) = \det(\mathbf{A}^\top \mathbf{C} \mathbf{A} - \mathbf{a}_r^\top c_r \mathbf{a}_r) = \det(\mathbf{A}^\top \mathbf{C} \mathbf{A})(1 - c_r \mathbf{a}_r (\mathbf{A}^\top \mathbf{C} \mathbf{A})^{-1} \mathbf{a}_r^\top) \\ = \det(\mathbf{A}^\top \mathbf{C} \mathbf{A}) R_{rr}. \quad (14)$$

Thus, the ratio of the determinant of the stiffness matrix of the modified and the initial system is identical to the redundancy of the element to be removed:

$$\frac{\det(\tilde{\mathbf{K}})}{\det(\mathbf{K})} = \frac{\det(\mathbf{A}^\top \mathbf{C} \mathbf{A} - \mathbf{a}_r^\top c_r \mathbf{a}_r)}{\det(\mathbf{A}^\top \mathbf{C} \mathbf{A})} = R_{rr}. \quad (15)$$

Equations (14) and (15) as well as the relation to the stiffness based robustness index proposed by Starossek and Haberland (2011) can be found in Gade (2024). It underlines the applicability of our approach of distributing redundancies homogeneously and by this maximizing the redundancy to achieve a robust structural design. The calculation procedure of the redundancy of an element can be made fast, offering an advantage regarding computational time compared to the procedure using the determinant (Tkachuk et al. 2023).

To showcase the above-mentioned approach of distributing the redundancy homogeneously within a structure, an optimization scheme using this objective is described in detail.

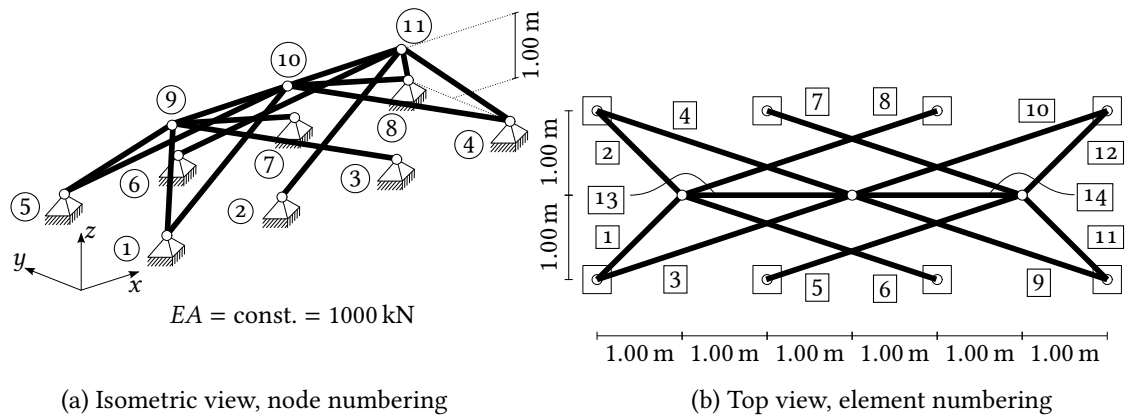

(a) Isometric view, node numbering  (b) Top view, element numbering

**Figure 3** Initial configuration of a 3D truss structure; Isometric view, node numbering and coordinate system shown in (a); Top view and element numbering shown in (b).

Figure 3(a) shows the initial configuration of a 3D truss structure in the isometric view with node numbering and the coordinate system. The top view and the element numbering can be seen in Figure 3(b). The structure consists of 14 truss elements with a constant element stiffness $EA = 1000$ kN and has a degree of static indeterminacy of $n_s = 5$. The spatial distribution of the redundancy is shown in color scheme in Figures 4(a) to 4(b). As it can be seen in the color scheme, the redundancy of the elements is varying between 0.08 and 0.58. The individual redundancies of the elements are additionally shown in Table 1 in line $R_{kk}$. The four elements at both ends of the structure drawn in dark blue have a very low redundancy and are of high importance for the load transfer. In case these elements fail, little possibilities for the redistribution of forces are given, thus these elements are very relevant for structural integrity.

In order to obtain a homogeneous distribution of the redundancies, the spatial locations of the nodal points are chosen as the design variables within the optimization. The optimization problem can then be formulated as follows:

$$\min_{\mathbf{s}} f(\mathbf{s}), \quad f(\mathbf{s}) = R_{\max} - R_{\min}, \quad \mathbf{s}^\top = \begin{bmatrix} x_1 & x_2 & x_9 & y_1 & y_2 & z_9 & z_{10} \end{bmatrix}. \quad (16)$$

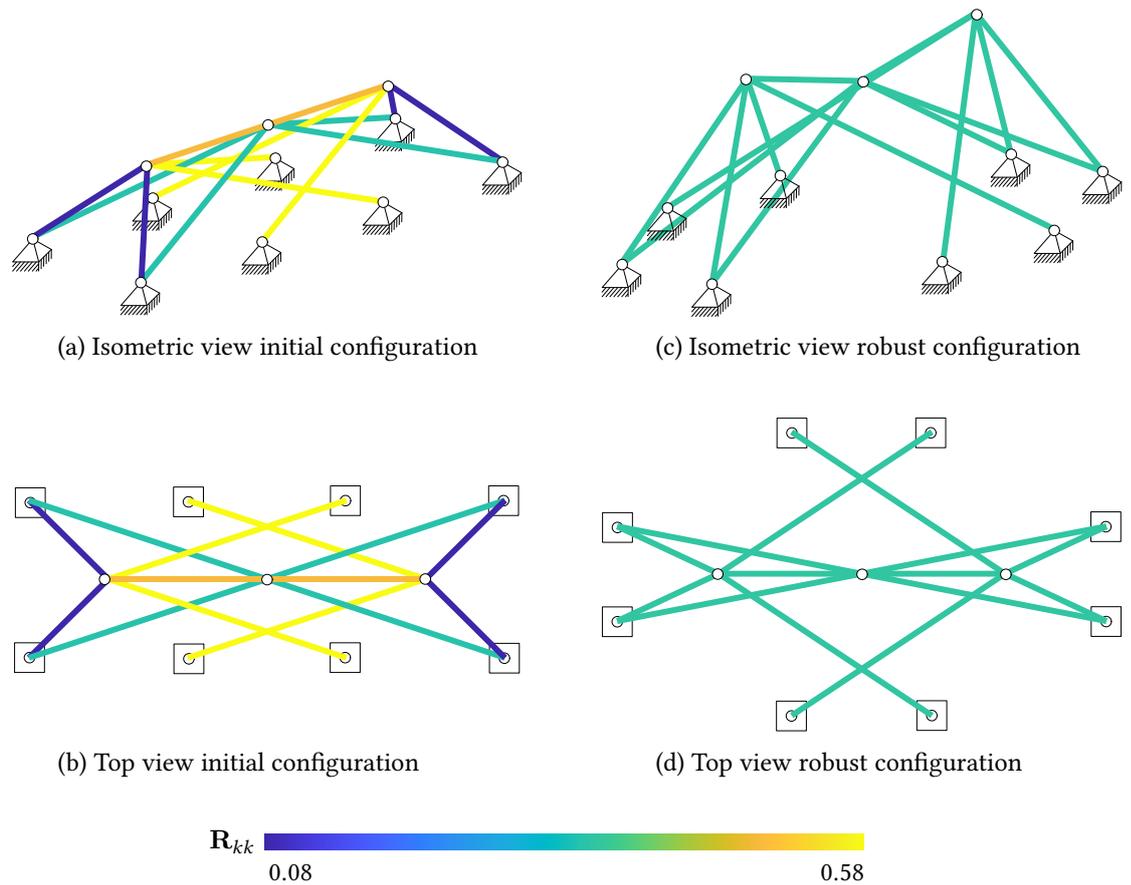

**Figure 4** Optimization of a 3D truss structure to obtain a robust design. Isometric view (a) and top view (b) of initial configuration shown on the left, colors indicating the redundancies according to the colorbar. Isometric view (c) and top view (d) of robust configuration shown on the right.

$R_{\max}$ and $R_{\min}$ denote the maximum and minimum redundancy of the structure, respectively. The remaining locations of the nodal points are chosen such that the structure remains symmetric and the support points do not move in $z$-direction, compared to the initial configuration. Therefore, only the seven values in Equation (16) are to be used as design variables within the optimization.

The optimization is performed with the commercial software MATLAB, using the sequential quadratic programming algorithm, as described in detail by Nocedal and Wright (2006). Figure 4(c) shows the optimized configuration in isometric view with the homogeneous redundancy distribution, as can be seen by the equal color of all elements. The top view of the optimized configuration is shown in Figure 4(d), clearly indicating the symmetry of the structure. The redundancies of the elements of the optimized configuration are shown in Table 1 in the row $R_{kk,\text{opt}}$.

This example shows that repositioning of nodes can be used to generate a structure with a homogeneous redundancy distribution. Finally, an exemplary study is performed to show that this structure is also more robust, *i.e.* yields smaller changes in element elongations due to a given arbitrary load and that the change in the determinant of the stiffness matrix is independent of the element to fail. The initial and robust configurations shown in Figure 4 are compared.

Table 2 shows in the second and third columns the changes in element elongations due to a load of 100 kN in vertical direction on nodes 9, 10 and 11. Each line refers to the structural system with one element removed, which is indicated in the first column. For most of these cases the change in element elongation for the robust system is significantly smaller compared to the initial system. But for certain elements, the increase in element elongation is smaller for the initial configuration, for example if element 5 is removed. This is in good accordance with the values of the redundancies, since for these elements the redundancy is large in the initial configuration and becomes smaller in the robust configuration. However, for the robust configuration, the changes in element elongation vary on a smaller scale and the arithmetic mean of the changes $\overline{|\Delta e|}$ is also smaller in comparison to the initial configuration. The same analysis could be done for any other given load or displacement and a similar result could be seen according to Equation (13).

| removed element $r$ | $\|\Delta e_r\|$ in μm | | $\det(\tilde{\mathbf{K}})/10^{23}$ in $(\text{kN m}^{-1})^9$ | | $\beta_r = \frac{\|\mathbf{d}_r\|_2 - \|\mathbf{d}\|_2}{\|\mathbf{d}\|_2}$ in % | |
|---|---|---|---|---|---|---|
| | init. config. | rob. config. | init. config. | rob. config. | init. config. | rob. config. |
| 1  | 1.34 | 0.15 | 0.95 | 3.89 | 96.83 | 11.06 |
| 2  | 1.34 | 0.15 | 0.95 | 3.89 | 96.83 | 11.06 |
| 3  | 0.52 | 0.31 | 3.97 | 3.89 | 78.84 | 116.12 |
| 4  | 0.52 | 0.31 | 3.97 | 3.89 | 78.84 | 116.12 |
| 5  | 0.08 | 0.19 | 6.64 | 3.89 | 2.29 | 22.99 |
| 6  | 0.08 | 0.19 | 6.64 | 3.89 | 2.29 | 22.99 |
| 7  | 0.08 | 0.19 | 6.64 | 3.89 | 2.29 | 22.99 |
| 8  | 0.08 | 0.19 | 6.64 | 3.89 | 2.29 | 22.99 |
| 9  | 0.52 | 0.31 | 3.97 | 3.89 | 78.84 | 116.12 |
| 10 | 0.52 | 0.31 | 3.97 | 3.89 | 78.84 | 116.12 |
| 11 | 1.34 | 0.15 | 0.95 | 3.89 | 96.83 | 11.06 |
| 12 | 1.34 | 0.15 | 0.95 | 3.89 | 96.83 | 11.06 |
| 13 | 0.04 | 0.01 | 5.64 | 3.89 | 0.60 | 0.18 |
| 14 | 0.04 | 0.01 | 5.64 | 3.89 | 0.60 | 0.18 |
| | $\overline{|\Delta\mathbf{e}|} = 0.56$ | $\overline{|\Delta\mathbf{e}|} = 0.18$ | | | $\overline{\beta} = 50.93$ | $\overline{\beta} = 42.93$ |

**Table 2** Changes in element elongation due to a prescribed load of 100 kN on nodes 9, 10 and 11 in vertical direction, determinant of modified stiffness matrix and relative change of the norm of the displacements due to the prescribed load. Different structural configurations representing initial configuration and robust configuration for an individual element's removal.

In columns four and five, Table 2 shows the determinant of the stiffness matrix of the system with one element removed. It can be seen that the determinant is independent of the element to fail for the robust configuration, for which the redundancies are distributed homogeneously. For the initial configuration, the changes can be compared to Equation (15) and the redundancy values in Table 1.

Additionally, the last two columns of Table 2 compare the initial and robust configuration with respect to the change in the Euclidean norm of the complete displacement vector. For both configurations, the displacements due to the aforementioned vertical load of 100 kN on the three top nodes are calculated for the intact system $\mathbf{d}$ and the system with one element removed $\mathbf{d}_r$. Each line in the table shows the relative change $\beta_r$ in the Euclidean norm of the displacement vectors for the case that one element is removed. For the removal of certain elements, the relative change $\beta_r$ is slightly larger for the robust configuration. But the arithmetic mean of all configurations shows that the robust configuration leads to less change in displacements in case of an element failure.

The assumptions of the optimized configuration being symmetric can of course also be neglected and various different solutions exist that satisfy the homogeneous redundancy distribution. Another approach to achieve this goal would be to use the cross-sections as design variables. In case of adjustments of the cross-sectional thickness in hollow sections, this makes the geometrical appearance independent of the optimization.

## 4 Assemblability

### 4.1 Imperfection Induced Strains

Regarding the construction of a structure, there are multiple factors that need to be taken into account for the assembly process. How can individual members be assembled without obstructing each other and how is the accessibility for cranes maintained? In case of a steel structure, are the connections bolted or welded, or how are the connections made? The question that we are addressing in this contribution is, how can imperfection induced strains due to manufacturing errors be considered in the planning stage? This aspect is particularly relevant for highly optimized structures which are especially sensitive to imperfections.

From a structural engineering point of view, one goal is to avoid large stresses induced during on-site assembly due to manufacturing imperfections of certain elements. Since the stresses are proportional to the strains for a constant Young's Modulus, strains will be used here to assess the structure with regard to the imperfection sensitivity.

The magnitude of the imperfections in length are assumed to be relative to the length of the respective element. As described in Section 2.2, for truss structures, each column $k$ of the redundancy matrix represents the negative elastic elongations in all elements that occur, if a prescribed unit elongation is applied on element $k$. Therefore, we can use the redundancy matrix scaled column-wise by the lengths of the elements to evaluate the strains induced by the imperfections. For this, we introduce the diagonal matrix $\mathbf{L} \in \mathbb{R}^{n_e \times n_e}$ that contains the lengths of the individual elements on the main diagonal:

$$\mathbf{L} = \begin{bmatrix} L_1 & 0 & \cdots & 0 \\ 0 & L_2 & \cdots & 0 \\ \vdots & \vdots & \ddots & \vdots \\ 0 & 0 & \cdots & L_{n_e} \end{bmatrix}. \tag{17}$$

Furthermore, the matrix $\boldsymbol{\alpha} \in \mathbb{R}^{n_e \times n_e}$ is introduced to specify the magnitude of the imperfection as the percentage of the original length for each element individually:

$$\boldsymbol{\alpha} = \begin{bmatrix} \alpha_1 & 0 & \cdots & 0 \\ 0 & \alpha_2 & \cdots & 0 \\ \vdots & \vdots & \ddots & \vdots \\ 0 & 0 & \cdots & \alpha_{n_e} \end{bmatrix}. \tag{18}$$

$\mathbf{E}_{\text{ass}} \in \mathbb{R}^{n_e \times n_e}$ expresses now column-wise the elastic elongations in all members caused by imperfections:

$$\mathbf{E}_{\text{ass}} = -\mathbf{R}\boldsymbol{\alpha}\mathbf{L}. \tag{19}$$

To obtain the strains in each element from these elongations, the entries of $\mathbf{E}_{\text{ass}}$ need to be divided row-wise by the original length of the respective element:

$$\boldsymbol{\varepsilon}_{\text{ass}} = \mathbf{L}^{-1}\mathbf{E}_{\text{ass}} = -\mathbf{L}^{-1}\mathbf{R}\boldsymbol{\alpha}\mathbf{L}. \tag{20}$$

The matrix $\boldsymbol{\varepsilon}_{\text{ass}} \in \mathbb{R}^{n_e \times n_e}$ contains column-wise the distribution of strains in the structure due to a length imperfection relative to the original length in one element. Compared to a standard finite element calculation of imperfection-induced strains, the above proposed procedure offers a compact matrix-based calculation that avoids repetitive analysis of the full structure. Different norms can now be applied to the columns $k$ to define a measure that can be compared easily. While the maximum norm $\max_i(\varepsilon_{\text{ass},ik})$ concentrates on the largest value of strain induced by an imperfection, the Euclidean norm $||\varepsilon_{\text{ass},ik}||_2$ takes into account the effect on all members of the structure. The effect of imperfections in the members of the structure can now be compared and the design and/or assembly sequence adapted accordingly. In order to evaluate the effect of all imperfections, a corresponding matrix norm can be applied to the complete matrix $\boldsymbol{\varepsilon}_{\text{ass}}$.

In the following, we will showcase the influence of manufacturing imperfections and how the influences can be altered within an optimization scheme. In a second example, different assembly sequences are compared with regard to intermediate strain states showing that the sequence itself is largely influencing the maximum strain throughout the construction process.

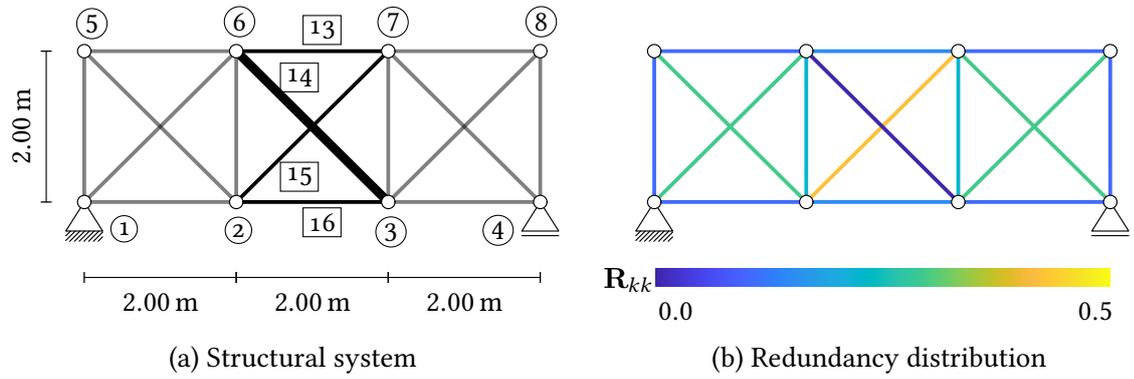

(a) Structural system  (b) Redundancy distribution

**Figure 5**  Truss structure with two prefabricated modules (grey) and four elements for final assembly (black); Element 14 with 100 times stiffness compared to all other elements (a). Redundancy distribution of the structure (b) in color scheme.

**Table 3**  Assessment of assembly parameters for initial truss structure (Figure 5).

| Element $k$ | 13 | 14 | 15 | 16 |
|---|---|---|---|---|
| $\mathbf{R}_{kk}$ | 0.1504 | 0.0043 | 0.4254 | 0.1504 |
| $\max_i(\varepsilon_{\text{ass},ik})$ | 0.0213 | 0.0425 | 0.0425 | 0.0213 |
| $\|\varepsilon_{\text{ass},ik}\|_2$ | 0.0361 | 0.0722 | 0.0722 | 0.0361 |

### 4.2 Influence of Geometric Imperfections

Figure 5 shows a simple 2D truss structure with node and element numbering on the left and the redundancy distribution in color scheme on the right. The stiffness of element 14 is 100 times higher than the constant stiffness of all other elements, and therefore the element is drawn thicker. This leads to a very low redundancy for element 14. The total degree of static indeterminacy is $n_s = 3$. In this scenario, the imperfection in length is defined as 10 % of the perfect length of the members, i.e., $\boldsymbol{\alpha} = 0.1 * \mathbf{1}_{n_e}$.

The grey elements on either side of the structure are assumed to be pre-fabricated and thus no geometric imperfections are assumed for them. The black elements 13 to 16 are used for final assembly on site. In this study we are interested in the influence of manufacturing imperfections for different elements. Element 14 is the one with the lowest redundancy, meaning that according to the interpretation of the redundancy matrix, it is the least constrained by the surrounding. Nevertheless, the maximum strain and the Euclidean norm of the strain is larger in comparison to the elements 13 and 16, see Table 3. This means that for the scenario that one element is imperfectly manufactured, element 13 or 16 would influence the strain distribution on a smaller scale in comparison to elements 14 and 15.

In a second scenario, where element 14 is said to be imperfectly manufactured, the strains that are induced by a length imperfection should be minimized. This can be done by a shape optimization using the nodal positions as design variables. It is prescribed that the supports remain at their original position, the lower chord of the truss remains straight and the system stays symmetric. Therefore, only 5 design variables are used and the locations of the remaining nodes are derived from these design variables. During the optimization, the Young's modulus and the cross-sections are kept constant. The optimization problem can be defined as follows:

$$\min_{\mathbf{s}} f(\mathbf{s}), \quad f(\mathbf{s}) = \|\varepsilon_{\text{ass},i14}\|_2, \quad \mathbf{s}^\top = \begin{bmatrix} x_2 & x_5 & x_6 & y_5 & y_6 \end{bmatrix} \tag{21}$$

The optimization was again performed with Matlab using sequential quadratic programming. Figure 6 shows the original configuration of the truss on the left and the optimized geometry according to Equation (21) on the right. Table 4 shows the resulting values for the redundancy, the maximum strain, and the Euclidean norm of the strain. One can see that the Euclidean norm of the strain was reduced by 12 % from 0.0722 to 0.0632. One can also see that the difference in the Euclidean norm between element 13 and 14 decreased drastically, meaning that the impact of the change in length regarding the strains decreased from 100 % difference in the initial configuration to 13 % in the optimized configuration.

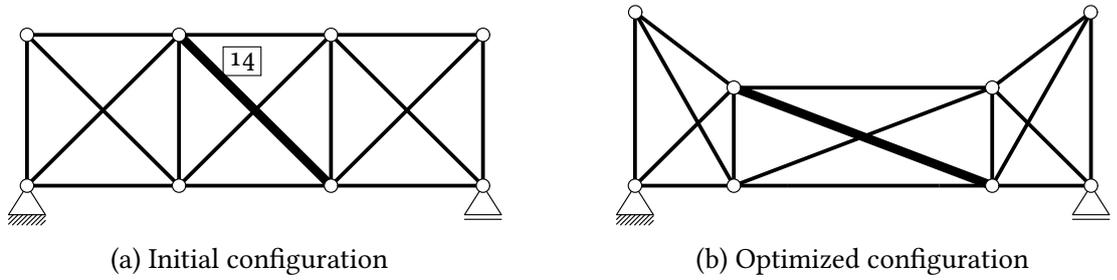

(a) Initial configuration            (b) Optimized configuration

**Figure 6** Truss structure from introductory example with stiffer diagonal element 14 (a). Structure with optimized nodal positions to minimize $||\varepsilon_{\mathrm{ass},i14}||_2$ (b).

**Table 4** Assessment of assembly parameters for optimized truss structure (Figure 6).

| Element $k$ | 13 | 14 | 15 | 16 |
|---|---|---|---|---|
| $\mathbf{R}_{kk}$ | 0.3001 | 0.0037 | 0.3682 | 0.3001 |
| $\max_i(\varepsilon_{\mathrm{ass},ik})$ | 0.0321 | 0.0368 | 0.0368 | 0.321 |
| $||\varepsilon_{\mathrm{ass},ik}||_2$ | 0.0551 | 0.0632 | 0.0632 | 0.0551 |

### 4.3 Assembly Sequence

The following case study aims to understand the influence of different assembly sequences on the strain state within a structure. For different states of the assembly $l$, reaching from the first assembled element $a$ to the last element $f_l$ assembled in this step, the strain distribution can be calculated in vector format as follows:

$$\varepsilon_{\mathrm{seq}}^l = \sum_{k=a}^{f_l} \varepsilon_{\mathrm{ass},ik}^l. \tag{22}$$

The matrix $\varepsilon_{\mathrm{ass}}^l$ describes the state $l$ within the assembly sequence. Since there exist many possibilities with various intermediate structural configurations for the assembly sequence, an efficient update can drastically decrease the computational effort (Krake et al. 2022).

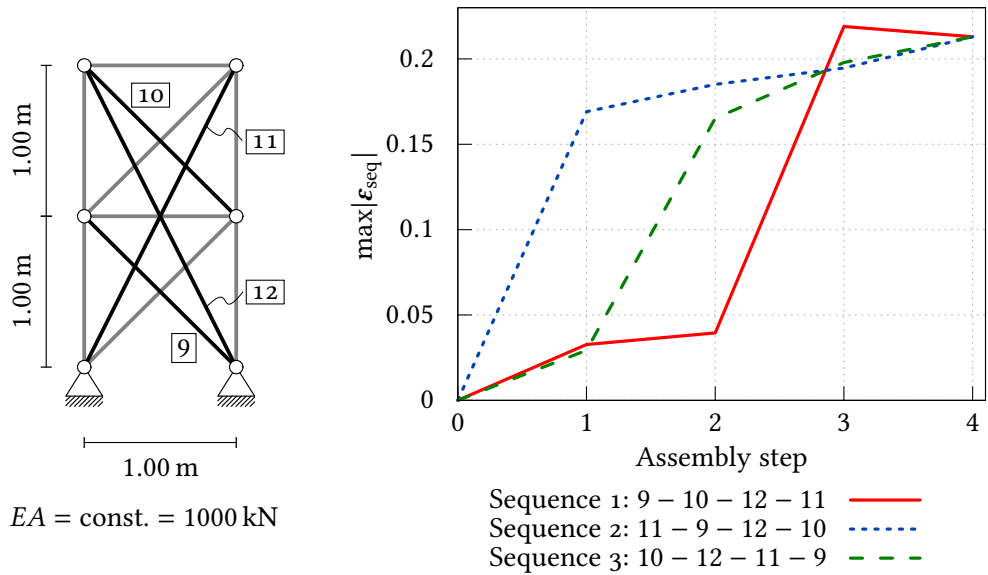

**Figure 7** Truss structure of already assembled elements in grey and four elements for final assembly in solid black (left). Maximum strain of all assembled elements throughout different assembly sequences (right). Sequence 1 outreaches the final maximum strain and proves therefore unfeasible.

Figure 7 shows the structural system of a plane truss on the left. The statically determinate part of the structure is shown in light grey and is said to be constructed without any geometric imperfections. The elements labelled 9 to 12, drawn in solid black, are the ones that will be assembled with given imperfections of $\alpha_9 = \alpha_{10} = 0.1$, $\alpha_{11} = 0.3$ and $\alpha_{12} = -0.3$. On the right of Figure 7, the maximum absolute strain of three exemplary construction sequences is shown with different colors. The x-axis represents the assembly steps, starting from the initial step 0 to the final assembly step 4. On the y-axis, the maximum absolute strain value of all elements of the structure is given, according to Equation (22). In the initial state, the strain is zero in the whole structure. In the final state, the maximum absolute value is similar for all assembly sequences. Since we are dealing with linear static analyses, the theorem of Betti-Maxwell applies and the final distribution of strains is independent of the assembly sequence.

One can obtain that different assembly sequences yield different maximum strains and thus stresses throughout the process. The sequence that is drawn in red, where element 11 is assembled last, yields a higher maximum strain at step 3 than in the final state. This means that sequence 1 should be avoided, otherwise the intermediate maximum strain outreaches the one that is unavoidable in the final assembly. One could of course also track the strain of individual elements throughout the assembly process to choose a sequence that is defined as optimal for any given scenario. This can be especially useful if a specific element is very sensitive to initial strains or if for example sensors are placed and initial strain deviations should therefore be avoided.

## 5  Conclusion

The paper addresses the assessment of structures using the redundancy matrix and, by this, using the distribution of static indeterminacy. On this basis, quantitative performance indicators for the robustness and assemblability are presented. These additional measures for structural assessment enlarge the possibilities for design exploration in very early design stages.

A detailed derivation of the matrix calculations for these two structural performance indicators was given and showcased with various examples. It was shown that the design of robust structures can be achieved by distributing the redundancy homogeneously within the structure. Different measures for the structural performance were used to compare the robustness of an initial configuration and an optimized configuration with a homogeneous redundancy distribution. In the context of the construction process, the influence of geometric imperfections and the assembly sequence on initial strains was predicted using the redundancy matrix. By optimizing the assembly sequence the maximum initial strain could be reduced.

The presented methods are applicable to truss and frame structures and can be especially useful in building systems that are sensitive towards imperfections. The extension of the notion of the redundancy matrix to plates and shells is ongoing work. In addition, the extension of the redundancy matrix to the non-linear setting is work in progress and will allow a straightforward transfer of the proposed indicators to non-linear problems as well.

The application of the presented methods to the behavior of a structure during progressive collapse is still an open question. In this situation, the redundancy matrix changes constantly after damage starts. In each damage state the presented methods can be applied, but a repeated update of the redundancy matrix is necessary.

**Authors' contributions**   DF: Conceptualization, Data curation, Formal Analysis, Investigation, Methodology, Software, Validation, Visualization, Writing – original draft, review & editing. MvS: Conceptualization, Investigation, Methodology, Validation, Project administration, Resources, Supervision, Writing – original draft, review & editing.

**Funding**   This work is partly funded by *Deutsche Forschungsgemeinschaft* (DFG, German Research Foundation) under Germany's Excellence Strategy – EXC 2120/1 – 390831618.

**Competing interests**   The authors declare that they have no competing interests.

**Journal's Note**   JTCAM remains neutral with regard to the content of the publication and institutional affiliations.